\documentclass[%
 reprint,
 superscriptaddress,
 amsmath,amssymb,
 aps,
 pra,
floatfix,
nofootinbib
]{revtex4-2}

\usepackage[T1]{fontenc}
\usepackage{mathptmx}
\usepackage{graphicx}
\usepackage{dcolumn}
\usepackage{physics}
\usepackage{bm}
\usepackage{times}
\usepackage{hyperref}
\hypersetup{
    colorlinks=true,
    linkcolor=blue,
    filecolor=blue,      
    urlcolor=blue,
    citecolor=blue
    }

\begin{document}

\title{Variations of the Kibble-Zurek scaling exponents of trapped Bose gases}

\author{Tenzin Rabga}\email{trabga@snu.ac.kr}
 \affiliation{Center for Correlated Electron Systems, Institute for Basic Science, Seoul 08826, Korea}
\author{Yangheon Lee}
 \affiliation{Center for Correlated Electron Systems, Institute for Basic Science, Seoul 08826, Korea}
 \affiliation{Department of Physics and Astronomy, Seoul National University, Seoul 08826, Korea}
  \author{Dalmin Bae}
 \affiliation{Center for Correlated Electron Systems, Institute for Basic Science, Seoul 08826, Korea}
 \affiliation{Department of Physics and Astronomy, Seoul National University, Seoul 08826, Korea}
 \author{Myeonghyeon Kim}
 \affiliation{Center for Correlated Electron Systems, Institute for Basic Science, Seoul 08826, Korea}
 \affiliation{Department of Physics and Astronomy, Seoul National University, Seoul 08826, Korea}
\author{Y. Shin}\email{yishin@snu.ac.kr}
 \affiliation{Center for Correlated Electron Systems, Institute for Basic Science, Seoul 08826, Korea}
 \affiliation{Department of Physics and Astronomy, Seoul National University, Seoul 08826, Korea}
 \affiliation{Institute of Applied Physics, Seoul National University, Seoul 08826, Korea}
 
\date{\today}

\begin{abstract}

We study the vortex nucleation dynamics in inhomogeneous atomic Bose gases quenched into a superfluid phase and investigate the dependence of the Kibble-Zurek (KZ) scaling exponent on the underlying trap configuration. For samples in a number of different inhomogeneous traps, we observe the characteristic power-law scaling of the vortex number with the thermal quench rate, as well as an enhanced vortex suppression in the outer regions with lower particle density, in agreement with the causality effect as encapsulated in the inhomogeneous Kibble-Zurek mechanism (IKZM). However, the measured KZ scaling exponents show significant differences from the theoretical estimates, and furthermore their trends as a function of the underlying trap configuration deviate from the IKZM prediction. We also investigate the early-time coarsening effect using a two-step quench protocol as proposed in a recent study and show that the interpretation of the measurement results without including the causality effect might be misleading. This paper provides a comprehensive study of vortex formation dynamics in quenched Bose gases confined in inhomogeneous trapping potentials and calls for a refined theoretical framework for quantitative understanding of the phase transition and defect formation processes in such inhomogeneous systems. 

\end{abstract}

\maketitle


\section{Introduction} 

The Kibble-Zurek mechanism (KZM) provides a framework for understanding the universal, non-equilibrium dynamics of a system undergoing a second-order phase transition~\cite{Kibble_1976,Zurek1985,ZUREK1996177}. Such a process is accompanied by the formation of independent phase domains as a system crosses the critical point of the transition. The merging of these domains leads to the formation of topological defects in the system. One of the ways in which this universality manifests itself is through the dependence of the defect number density on the rate at which the system is quenched through the critical point~\cite{del_Campo_2013,Dziarmaga10}. The KZM predicts a power-law scaling of the defect number density with the sample quench rate, as observed in many experimental studies using superfluid helium~\cite{Hendry1994,Bauerle1996,Ruutu1996,Dodd98}, liquid crystal~\cite{Chuang91}, superconductors~\cite{Carmi00,Monaco02}, ion chains~\cite{Pyka2013,Ulm2013,Ejtemaee13}, and ultracold atomic gases~\cite{Sadler2006,Weiler2008,Lamporesi2013,Goo21,Navon15,Ko2019,Donadello16}.

The original predictions for the Kibble-Zurek (KZ) scaling exponent consider homogeneous systems in which the entire sample simultaneously undergoes phase transition. Ultracold atomic gas experiments often use inhomogeneous trapping potentials for confining and cooling the thermal sample. This leads to inhomogeneous particle density distributions. Therefore, although the defect number density displays a power-law dependence on the quench rate, a quantitative comparison with theory necessitates that the inhomogeneity of the system be taken into account.

Spontaneous defect formations in inhomogeneous and homogeneous systems differ in two crucial respects. First, in inhomogeneous samples, not only the critical temperature but also the relative thermal quench rate acquires a local dependence, with different parts of the system undergoing phase transition at different times. Second, the defect formation is causally determined by the competition between the spread of the local order information and the speed of propagation of the phase transition front. When the phase transition front propagates significantly slowly, the corresponding slow growth of the ordered phase may lead to defect suppression. We call this the causality effect~\cite{kim22}. Such considerations lead to the inhomogeneous Kibble-Zurek mechanism (IKZM) which accounts for the phase transition dynamics in systems with density inhomogeneities~\cite{del_Campo_2011}, and in power-law trapping potentials with appropriately chosen quench rates, it predicts a larger scaling exponent for a system with the same universality class as the analogous homogeneous case.

Recent developments in large-area sample production in ultracold atomic gas experiments have enabled in-depth studies of the KZM, where the normal-to-superfluid phase transition is driven by controlled evaporative cooling and quantum vortices are created as effective topological point defects. Using strongly interacting $^{6}$Li Fermi gases, the universal power-law scaling behavior was demonstrated in the BCS$-$Bose-Einstein condensate (BEC) crossover regime~\cite{Ko2019}, where the microscopic character of the pairing mechanism changes from loosely bound Cooper pairs to tightly bound molecules. In subsequent experiments with large-area $^{87}$Rb samples, saturation of defect numbers for rapid quenches was observed and investigated. Such saturation behavior, which lies beyond the conventional KZ scenario, has revealed the possible significance of early-time coarsening of the order parameter during the initial stages of condensate growth~\cite{Goo21}. Moreover, the large-area samples allow for statistical analysis of the defect position distribution within the quenched, inhomogeneous samples, where an increase of the KZ scaling exponent was observed in the outer regions with higher atomic density gradient, compared to the central region~\cite{kim22}. This qualitatively corroborates the causality effect.

Although a power-law scaling of the defect number within a relatively slow quench regime was observed in all of these experiments, a quantitative understanding of the measured KZ scaling exponent for inhomogeneous atomic gases has not been fully established yet. In the study with strongly-interacting inhomogeneous Fermi gases of $^{6}$Li in a harmonic trapping potential, the measured KZ scaling exponent ($\alpha^{}_{\text{KZ}} = 2.24(9)$) showed a decent agreement with the theoretical estimate of $\approx 7/3$ from IKZM~\cite{Ko2019}. However, in an elongated sample of Bose gas of $^{87}$Rb, with a quartic density profile along the axial direction~\cite{Lim21}, the measured value of the KZ scaling exponent ($\alpha^{}_{\text{KZ}} = 2.6(1)$) was observed to be significantly larger than its estimated value of $\approx 16/9$ under the power-law trap assumption~\cite{Goo21}. This discrepancy necessitated a study of the trap geometry dependence of the different mechanisms involved in the vortex formation process. In addition, a subsequent study examining the role of the above-mentioned early-coarsening effect in vortex number suppression highlighted its implications not only for the saturation of vortex numbers at high quench rates but also for the resultant changes to the KZ scaling exponent~\cite{Gooetal22}.

As a natural extension, in this paper, we present a study of spontaneous defect formation in a number of inhomogeneous samples of quenched Bose gas of $^{87}$Rb. In a range of different trap configurations, we observe the characteristic power-law scaling of the defect number with the thermal quench rate, as well as the universal saturation behavior of vortex numbers at rapid quenches. However, we find that the measured values of the scaling exponents $\alpha^{}_{\text{KZ}}$, as well as their trends as a function of the trap configuration, significantly differ from the theoretical estimates based on the IKZM. This is in contrast to our analysis of the position dependence of the quench dynamics, which supports the causality effect in IKZM~\cite{kim22}. Subsequently, by adopting the two-step quench protocol~\cite{Gooetal22}, we extend the study to investigate the effect of early-time coarsening of the order parameter on vortex suppression. In a harmonic trap, we observe that the vortex suppression factor also exhibits a power-law dependence on the quench rate after passing the critical temperature, but surprisingly, this scaling exponent is comparable to the measured $\alpha_\text{KZ}$. This indicates the limitation of our current treatment of the two-step quench measurement results for inhomogeneous samples, where the causal interactions at the boundary of the initial seed condensate are assumed to be insignificant. 
 
Overall, this paper examines various aspects of the phase transition process and their dependence on the underlying trap configurations, and emphasizes the need for a refined theoretical framework for a quantitative understanding of the observed variations of the KZ scaling exponents.

The rest of the paper is organized as follows. We first describe our experimental setup and the different trap configurations in Sec.~II. In Sec.~III, we present and discuss the measurement results of the power-law scaling and saturation of the vortex number, the position-dependent suppression effect, and the two-step quench experiment. Sec.~IV provides some concluding remarks.

\section{Experimental Setup}

In order to study the effect of the trap geometry on the phase transition dynamics, we exploit the tunability of the clipped Gaussian optical dipole trap (ODT), where a slit with an adjustable width is used for truncating the incident Gaussian ODT beam and modifying the resultant oblate ODT profile. The use of this slit for generating a large-area BEC in a clipped Gaussian ODT was described in Ref.~\cite{Lim21}. As the original Gaussian ODT is successively clipped, the axial trapping potential changes from a harmonic trap to a trap with a quartic confinement. We also note that the elongation along the axial direction due to clipping results in an increased aspect ratio of the trap.

\begin{figure}[t]
    \centering
    \includegraphics[width=3.0in]{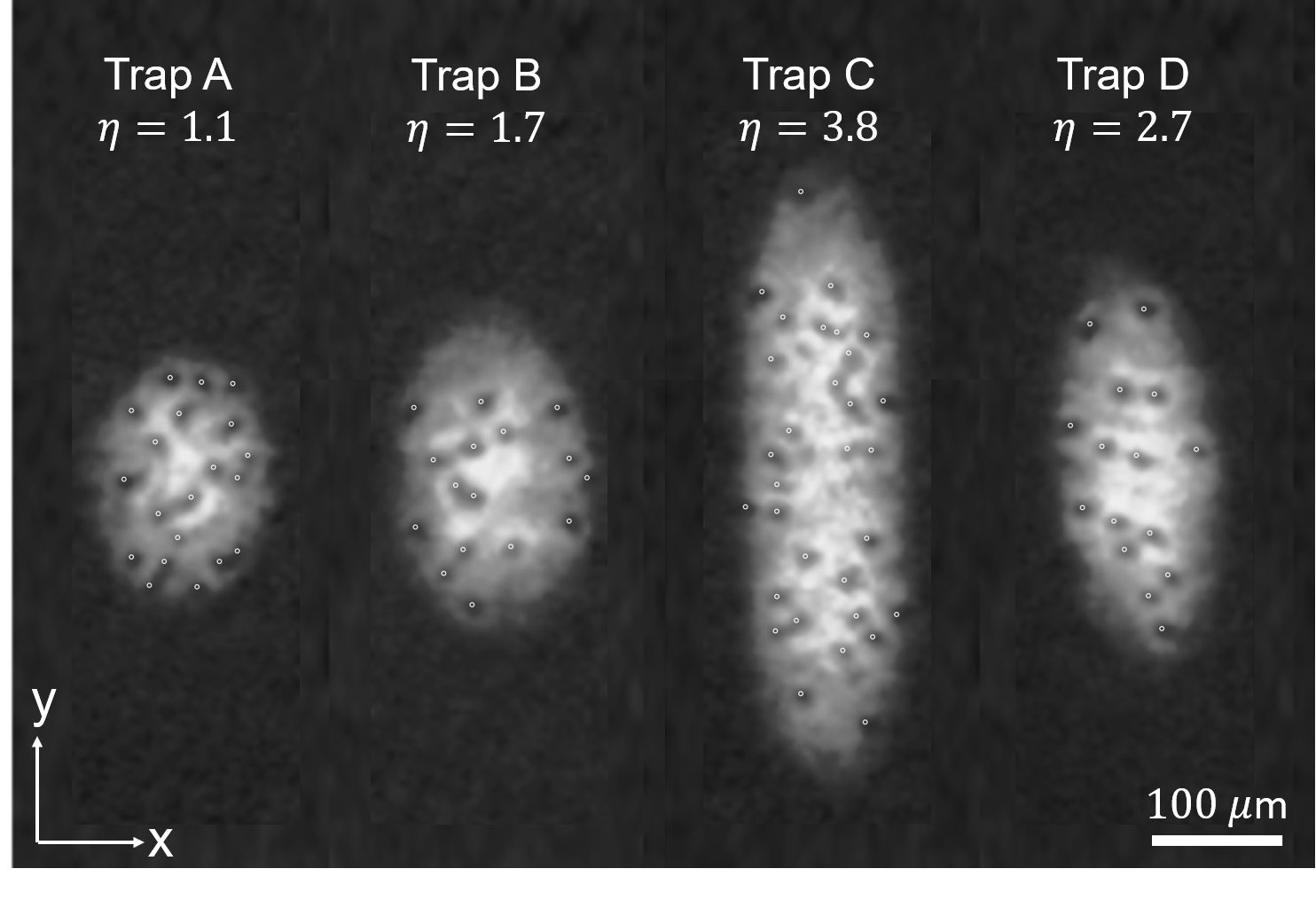}%
    \caption{Inhomogeneous BECs in four different trapping potentials, referred to as trap A, B, C, and D, respectively. $\eta = R_{y}/R_{x}$ denotes the sample aspect ratio, with $R_{x(y)}$ being the Thomas-Fermi radius of the trapped condensate for the $x(y)$ direction. Traps B and C are obtained by focusing a truncated Gaussian laser beam~\cite{Lim21}. The images were taken after a time of flight, showing BECs produced in the respective traps along with quantum vortices marked by white dots.}
    \label{fig:trapconfigs}
\end{figure}

In this paper, we present experimental results from four different trap configurations (referred to as trap A, B, C, and D, respectively) that result in the condensate samples shown in Fig.~\ref{fig:trapconfigs}. For a given trap configuration, the experiment begins with a thermal sample of Rb atoms trapped in a 1070 nm ODT. The details of the sample preparation are similar to what was described in Ref.~\cite{Goo21}. The sample is evaporatively cooled by linearly lowering the trap depth from $U_{i} > U_{c}$ to $U_{f} < U_{c}$ in a variable time $t_{q}$, where $U_{c}$ is the critical trap depth at which a BEC emerges in the system. While we set $U_{i} = 1.15U_{c}$, $U_{f}$ ranges from $0.27U_{c}$ to $0.36U_{c}$ for the four traps. We define a quench rate $r_q$ characterizing how fast the system passes through the critical point as $r_q=\frac{U_i-U_f}{U_c}\frac{1}{t_q}$. After the sample is quenched to its final temperature, it is held for a hold time of $t_{h} = 1.25\, \text{s}$ in order to facilitate both the growth of the condensate fraction and the formation of vortices. The condensate is then imaged after a time-of-flight time of $40.9$ ms. 

Except for the degrees to which the ODT beam is clipped by the adjustable slit, traps A, B, and C share the same optical setup. Trap B is obtained by slightly clipping the Gaussian beam, while trap C is created by clipping the beam even further until a maximally elongated condensate is obtained. Trap C is identical to the trap used in our previous works in Ref.~\cite{Goo21,Gooetal22}. The traps are also labeled by the aspect ratio of the condensate, $\eta = R_{y}/R_{x}$, where $R_{x(y)}$ is the Thomas-Fermi radius along the \textit{x}(\textit{y}) direction. In order to explore the role of the trap aspect ratio on the defect formation dynamics, we include trap D in our paper. It is created by focusing an unclipped Gaussian beam and has the harmonic axial confinement as trap A, but with a much larger aspect ratio. The beam waist along the $z$ direction at the focus of the ODT beam, and therefore the trap center, is 25$\%$ larger for trap D compared to trap A, resulting in a slightly weaker trap and consequently a larger condensate.  

\begin{table}[t]
\caption{\label{tab:trapparams} Typical trap parameters for the four trap configurations used in this paper. The trapping frequencies $\omega_{x,y,z}$, the Thomas-Fermi radii $R_{x,y,z}$, and the condensate healing length $\xi$ at the trap center are shown.}

\begin{ruledtabular} 
\begin{tabular}{llll} Trap Config.  & $\omega_{x,y,z}/2\pi\, (\text{Hz})$ & $R_{x,y,z}\, (\mu\text{m})$ & $\xi\, (\mu\text{m})$\\ 
\hline 
~~~~~~~A & 7.6, 5.3, 225 & 76, 87, 2.0 & 0.20\\ 
~~~~~~~B & 7.8, N/A, 197 & 75, 129, 2.9 & 0.21\\ 
~~~~~~~C & 6.8, N/A, 164 & 65, 244, 2.6 & 0.27\\ 
~~~~~~~D & 8.2, 3.2, 85 & 57, 153, 5.5 & 0.25
\end{tabular} 
\end{ruledtabular} 
\end{table}

Table~\ref{tab:trapparams} shows the relevant trap parameters and typical sample conditions for the four different trap configurations. We note that although the trap is less tightly confined along the \textit{z} direction in trap D compared to the other traps, we assume that its is still well within the quasi-two-dimensional (2D) regime, as suggested by the appearance of highly visible vortex cores over the entire sample area (Fig.~\ref{fig:trapconfigs}). However, even in the case of a three-dimensionla (3D) sample with vortex line defects, the ensuing discussions about the results of experiments in this trap configuration still hold. Since the KZ power-law scaling exponent depends on the difference between the dimensions of the system $D$ and the defect $d$, a 2D sample ($D = 2$) with vortices ($d = 0$) has the same $D-d$ value as a 3D sample ($D = 3$) with vortex lines ($d = 1$), and therefore the same KZ scaling exponent~\cite{del_Campo_14}. 

As is seen in Fig.~\ref{fig:trapconfigs}, we observe multiple quantum vortices in the condensate images. For fast quenches, the maximum vortex numbers in each of the trap configurations easily exceed 20. In order to reduce the time taken to count these vortices and to accurately record them, we rely on a machine-learning-based vortex counting algorithm developed and optimized for our system, and adapted from Ref.~\cite{Metz21}.

\section{Results and Discussion}

\subsection{Power-law scaling and saturation} 

\begin{figure*}[t]
    \includegraphics[width=7in]{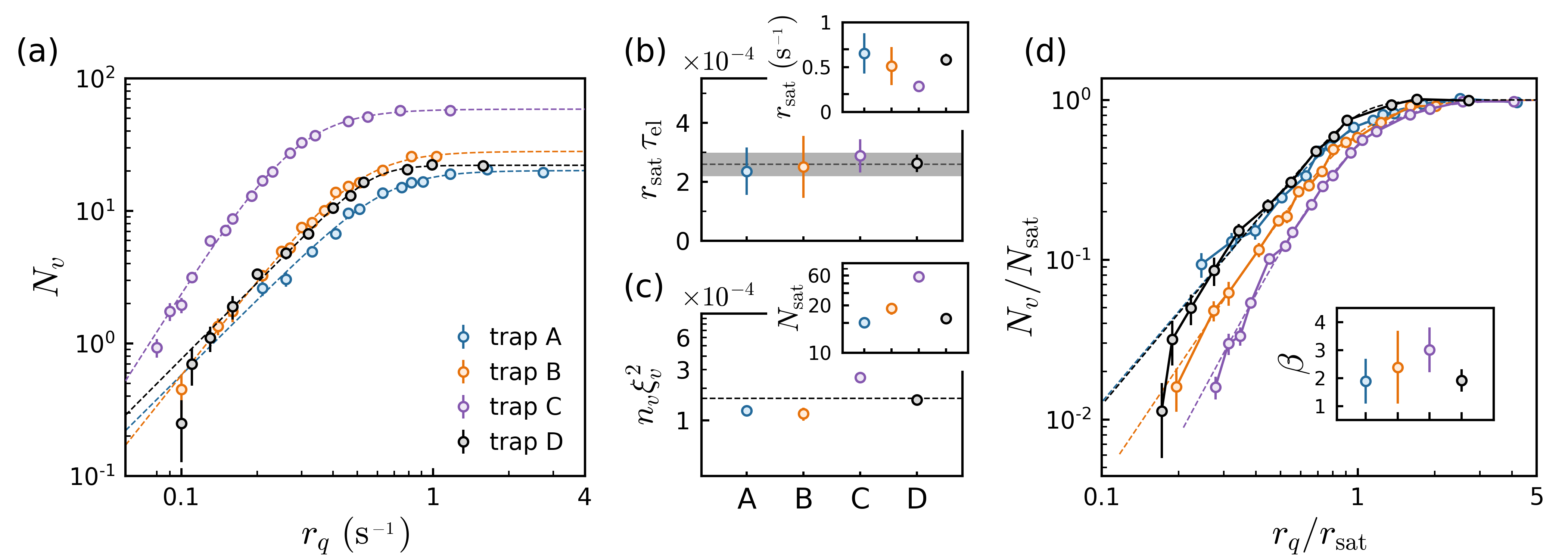}%

    \caption{Spontaneous vortex nucleation in inhomogeneous Bose gases quenched into a superfluid phase. (a) Log-log plot of the vortex numbers $N_{v}$ as a function of the quench rate $r_{q}$ for the four different traps A to D. Each data point is the average of 20 measurements and its error bar represents the standard error of the mean. The dashed lines are saturation-model fits to the data used for extracting the saturation quench rate $r_{\text{sat}}$ and saturated vortex number $N_{\text{sat}}$ (see the text for details and the insets in (b) and (c), respectively). (b) Normalized saturation quench rates $r_\text{sat} \tau_\text{el}$ for the four different trap configurations, where $\tau_{\text{el}}$ is the respective elastic collision time of the samples at critical temperature. The dashed line is the mean of the four data points and the gray band indicates the $1\sigma$ uncertainty bound. (c) Normalized saturated vortex densities $n_v \xi_v^2$, where $n_v=N_{\text{sat}}/(\pi R_x R_y)$ is the saturated vortex density and $\xi_v$ is the characteristic vortex core size. The dashed line indicates the mean value of the four data points. (d) $N_{v}/N_{\text{sat}}$ as a function of $r_{q}/r_{\text{sat}}$. The dashed lines show the same saturation-model fits as in (a). Inset: $\beta$ value for the different trap configurations.}
    \label{fig:allquench}%
\end{figure*}

In Fig.~\ref{fig:allquench}(a), we show the measurement results of the defect number as a function of the quench rate for the four different trap configurations. A couple of key observations can be made. First, we see that the vortex number exhibits a power-law scaling with the quench rate in a slow quench regime for all the trap configurations. Second, as the quench rate increases, the vortex number shows the characteristic saturation behavior, as was reported previously~\cite{Goo21,Ko2019,Donadello16}. To characterize the power-law scaling and saturation behavior of the defect number, we fit a phenomenological model to our data of the form $N_{v} = N_{\text{sat}}\left[1+\left(r_q/r_{\text{sat}}\right)^{-\beta \delta}\right]^{-1/\delta}$~\cite{Donadello16}. This allows us to extract the saturation vortex number $N_{\text{sat}}$, the saturation quench rate $r_{\text{sat}}$, and the scaling exponent $\beta$. The $\delta$ parameter describes the saturation behavior. 

In the inset of Fig.~\ref{fig:allquench}(b), we plot the saturation quench rates $r_{\text{sat}}$ obtained for each of the trap configurations. In our previous studies, defect saturation has been associated with the finite time required for a condensate to grow in the system~\cite{Goo21,Gooetal22,wolswijk22}. The elastic collision time at the critical temperature, $\tau_{\text{el}}$~\footnote{The elastic collision time is given by $\tau_{\text{el}} = (n\sigma \bar{v})^{-1}$, where $n$ is the particle density, $\sigma = 8\pi a^2$ is the collisional cross section with an \textit{s}-wave scattering length of $a$, and $\bar{v} = 4\sqrt{k_{B}T_{c}/\pi m}$ is the mean particle velocity.}, provides a characteristic time scale of the system's dynamics. Fig.~\ref{fig:allquench}(b) shows the saturation quench rates multiplied by the respective elastic collision times $\tau_{\text{el}}$ for the condensate samples at their respective critical temperatures $T_{c}$. We note that the normalized saturation rates for the different trap configurations are consistent with each other, suggesting a universal value of $r_{\text{sat}} \tau_{\text{el}} = 2.6(4)\times10^{-4}$, as indicated by the dashed line. This highlights the significance of universal dynamics of quenched Bose gases in understanding the defect saturation behavior.

As another probe for examining the universality of defect saturation, we consider the normalized saturated vortex density $\tilde{n}_v=n_{v}\xi_{v}^{2}$~\cite{Ko2019,Goo21}, where $n_{v} = N_{\text{sat}}/(\pi R_{x}R_{y})$ is the saturated vortex density and $\xi_{v} = \hbar/m c_{s}$ gives the average vortex core size, where $c_{s} = \sqrt{2\mu/3m}$ is the speed of sound in a condensate of oblate geometry with a chemical potential $\mu$ and particle mass $m$. As shown in Fig.~\ref{fig:allquench}(c), $\tilde{n}_v$ is consistently obtained to be in the range of $1 - 2.5 \times10^{-4}$ for our trap configurations, with a mean value of $1.6\times10^{-4}$, indicated by the dashed line. Compared to a Fermi gas confined in a harmonic trap~\cite{Ko2019}, the normalized saturated vortex density is roughly a factor of 2 larger in a Bose gas confined in a similar trapping potential geometry. The inset shows the corresponding $N_{\text{sat}}$ values for each trap configuration.

In Fig.~\ref{fig:allquench}(d), we plot the normalized vortex number as a function of the normalized quench rate, where the two axes are scaled by the saturated vortex number $N_{\text{sat}}$ and the saturation quench rate $r_{\text{sat}}$ of the individual traps, respectively. The dashed lines indicate the normalized saturation-model fits to the data. They clearly show the variations in the KZ scaling behavior for the four different trap configurations, which is the main subject of this paper. As indicated earlier, the $\beta$ parameter in the saturation-model fit also characterizes, at least qualitatively, the power-law scaling observed in these data. As shown in the inset, the trend in the central value of this parameter with respect to the trap configuration, further highlights the trap dependence of the scaling behavior.

\subsection{KZ scaling exponents}

While the $\beta$ parameter, extracted from the saturation-model fit, is suggestive of the power-law scaling behavior, in order to extract the scaling exponent $\alpha^{}_{\text{KZ}}$ reliably, it is necessary to properly determine the range of $r_q$ within which KZ scaling is valid. Evidently, the fast quench regime should be excluded due to vortex number saturation. The extremely slow quench regime should also be excluded due to the ``finite-size effect" which results in vortex number suppression~\cite{Pyka2013,del_Campo_2011}. Here we outline the method used in this paper for determining the appropriate KZ scaling regime. 

\begin{figure}[t]
    
    \includegraphics[width=3.2in]{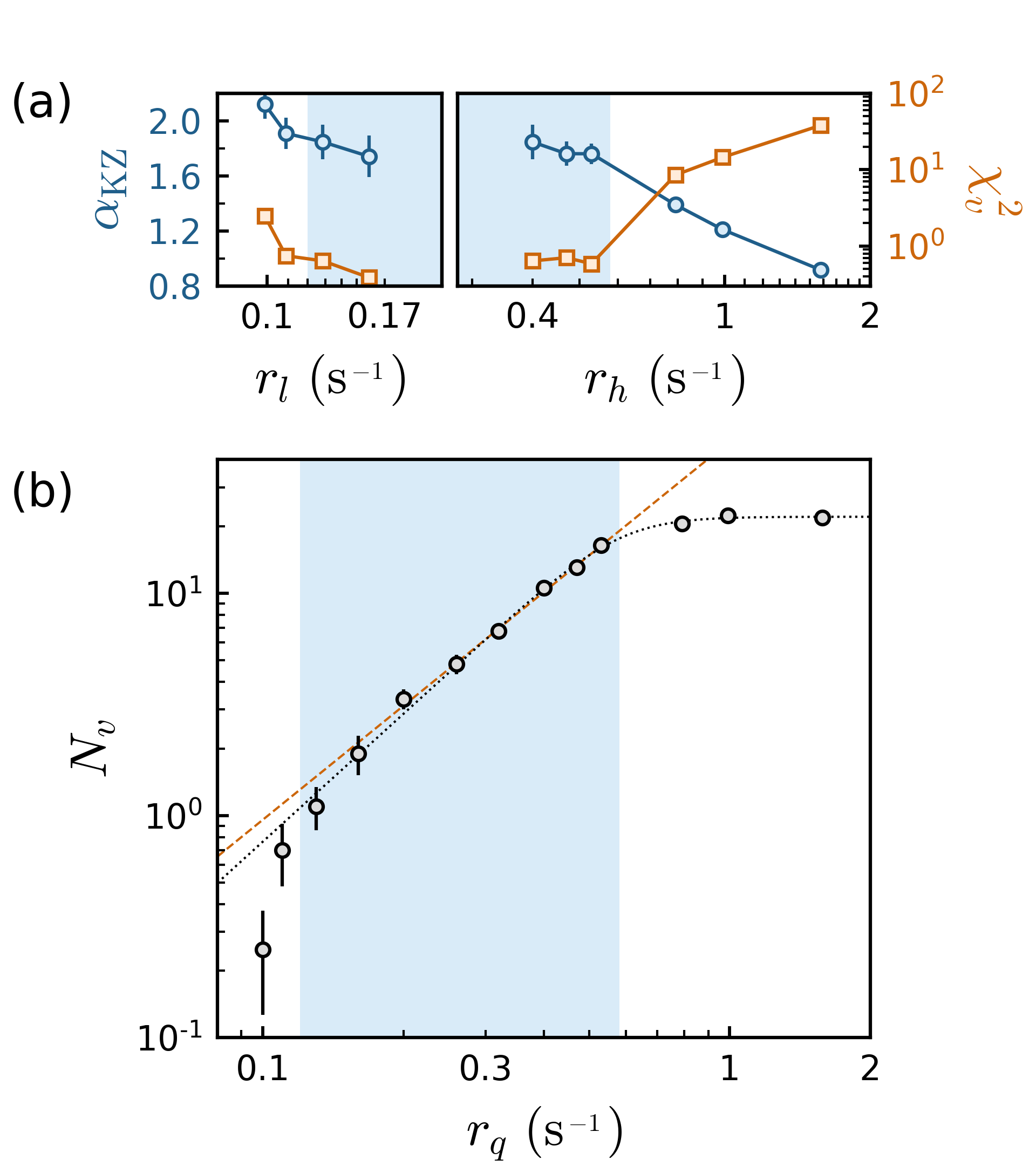}%

    \caption{KZ scaling region. The KZ scaling exponent $\alpha_\text{KZ}$ was determined from a power-law fit to the data in a restricted region given by $r_l<r_q<r_h$. (a) $\alpha^{}_{\text{KZ}}$ from the data of trap D for various fitting regions with (left) $r_h = 0.4\,\text{s}^{-1}$ and (right) $r_l = 0.12\,\text{s}^{-1}$. The corresponding $\chi^{2}_{\nu}$ values for the fits are also shown. (b) Vortex number measurement results for trap D [the same data as in Fig.~\ref{fig:allquench}(a)]. The blue-shaded region indicates the KZ scaling region suggested by the observed trends in $\alpha^{}_{\text{KZ}}$ and the corresponding values of $\chi^{2}_{\nu}$ in (a). The dashed line is the KZ power-law fit to the scaling region. The dotted black line is a saturation-model fit to the whole data.}
    \label{fig:akzchi2vscutoffs}
\end{figure}

To obtain the proper scaling region, we fit a power-law function of the form $N_{v} = N_{0}r_q^{\alpha^{}_{\text{KZ}}}$ to various subsets of the entire dataset and study the variations in the values of the $\chi^{2}_{\nu}$ of the fit and the corresponding $\alpha^{}_{\text{KZ}}$. The results of the vortex quench data obtained in our sample in harmonic trap D are shown in Fig.~\ref{fig:akzchi2vscutoffs}(b). The subset of the data that minimizes the $\chi^{2}_{\nu}$ and results in a stable value of $\alpha^{}_{\text{KZ}}$ is chosen as the appropriate KZ scaling region, indicated by the blue-shaded region. In Fig.~\ref{fig:akzchi2vscutoffs}(a), the left subfigure shows the effect of including low rate data points into our fit on the values of the scaling exponent $\alpha^{}_{\text{KZ}}$ and the fit $\chi^{2}_{\nu}$. The right subfigure shows the corresponding effects of including high rate data points into our fit on these values. As indicated by the plot of the $\chi^{2}_{\nu}$ value for different low and high rate cutoffs, excluding the two lowest and the three highest rate data points from our KZ scaling regime is justified as evidenced by the diverging values of the $\chi^{2}_{\nu}$ and the resultant values of $\alpha^{}_{\text{KZ}}$. With this choice of the KZ scaling region, indicated by the blue-shaded region in Fig.~\ref{fig:akzchi2vscutoffs}(b), we obtained a scaling exponent of $\alpha^{}_{\text{KZ}} = 1.8(1)$. The KZ scaling regions for all the other trap configurations are determined using the same strategy. Using this method, we observed that slow quench rates with $N_v<1$ were excluded from the fit region. This is not surprising since the finite-size effect becomes significant in the slow quench regime. We also find that the optimal high rate cutoff $r_{h}$ is approximately  $0.84r_{\text{sat}}$. Alternatively, this allows us to approximate the KZ scaling regime by setting the low cutoff rate $r_l$ to the lowest rate with $N_v>1$, and the high cutoff rate to $r_h = 0.84r_{\text{sat}}$.

\begin{figure}[t]
    \includegraphics[width=3.2in]{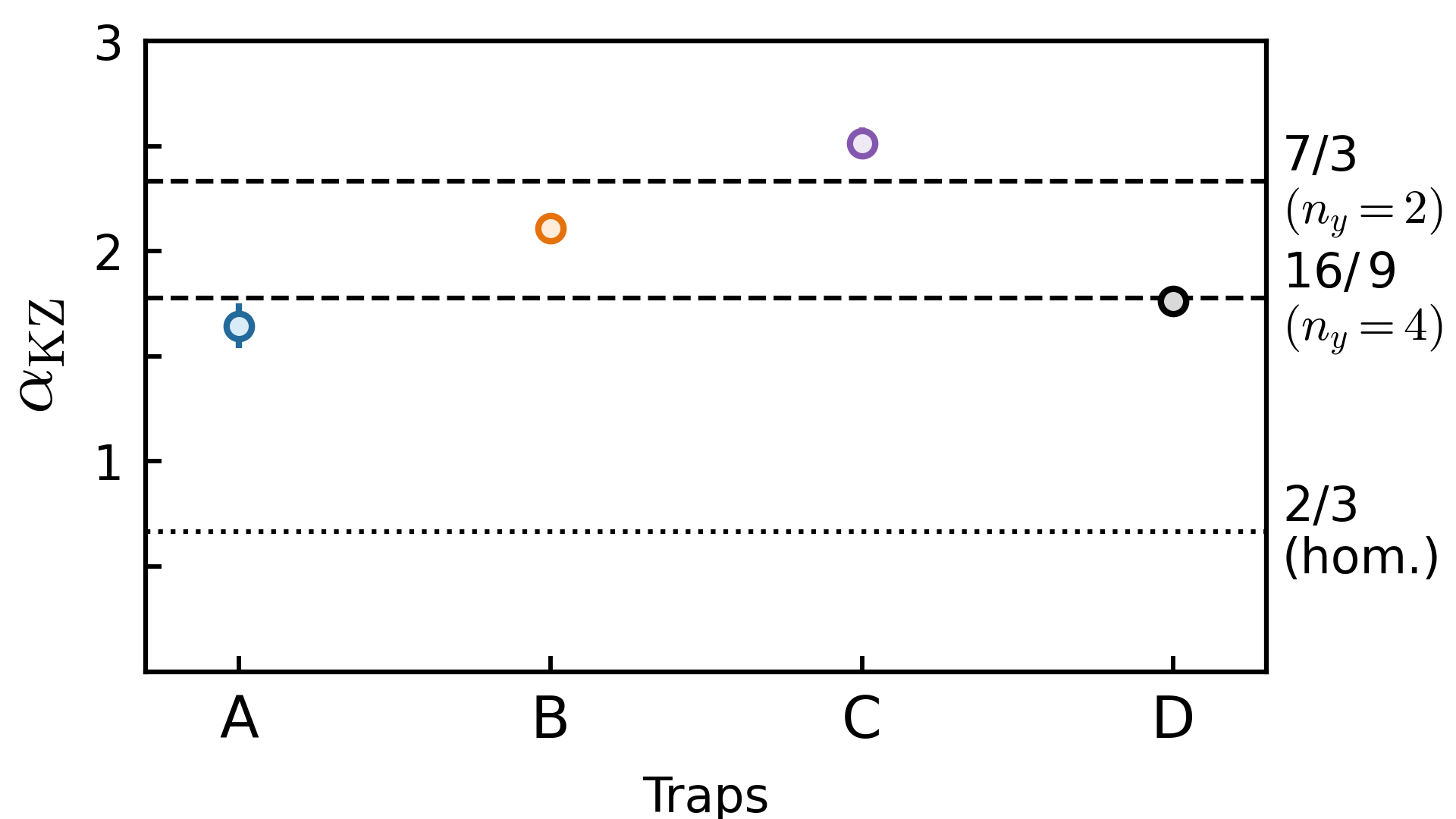}%
    \caption{KZ scaling exponents $\alpha^{}_{\text{KZ}}$ of inhomogeneous Bose gases for different trapping potentials. The error bars indicate the uncertainties obtained from the fit to the KZ scaling region for each trap configuration, determined using the scheme described in Fig.~\ref{fig:akzchi2vscutoffs}. Two horizontal dashed lines show the estimates from IKZM for samples trapped in power-law potentials with $\{n_x,n_{y}\} = \{2,2\}$ and $\{2,4\}$, respectively. The dotted line indicates the value of the KZ scaling exponent predicted for a homogeneous system.}
    \label{fig:aKZvstrap}
\end{figure}

Fig.~\ref{fig:aKZvstrap} displays the KZ scaling exponent as measured in the four different trap configurations. As observed in traps B and C, as trap A is elongated along the axial (\textit{y}) direction by clipping the incident Gaussian beam, the value of the scaling exponent increases. We also observe that despite the larger condensate size in trap D, the measured scaling exponent is consistent with what is measured in trap A, which is also a harmonic trap. We also note the similar trend displayed by the $\beta$ values for the saturation-model fits to the data for the different trap configurations, as shown in Fig.~\ref{fig:allquench}(d).

Within the framework of IKZM, some of these observations are quite surprising. As noted earlier, as an important consequence of the density inhomogeneity of the sample, different parts of the system undergo phase transition at different times. The probability of vortex formation is governed by the competition between the speed of the spread of the phase transition front and the local speed of sound. As a result, one should expect greater suppression of vortex formation in a more inhomogeneous sample, and therefore a decrease in the scaling exponent as the trap becomes more flattened. In other words, as the trap configuration changes from trap A to trap C, $\alpha^{}_{\text{KZ}}$ should approach the KZ scaling exponent for a 2D homogeneous sample given by $\alpha^{}_{\text{KZ,Hom}} = 2/3$ and indicated by a dotted line in Fig.~\ref{fig:aKZvstrap}. However, the measured values of the scaling exponents do not agree with this expectation.  

According to the IKZM, assuming a power-law trapping potential of the form $V(x,y) \propto |x|^{n_{x}} + |y|^{n_{y}}$, the scaling exponent is expected to have the following dependence on the trap parameters: $\alpha^{}_{\text{KZ}} = \frac{\left(\frac{1}{n_{x} - 1}+\frac{1}{n_{y}-1}\right)(1+\nu)+2\nu}{1+\nu z}$, where $\nu$ and $z$ are the static and dynamic critical exponents of the phase transition, respectively~\cite{del_Campo_2011}. With $\nu=2/3$~\cite{Donner07} and $z=3/2$~\cite{Navon15}, this predicts for $\{n_x,n_y\}=\{2,2\}$ (trap A and D) $\alpha_\text{KZ}=7/3$, and for $\{n_x,n_y\}=\{2,4\}$ (trap C) $\alpha_\text{KZ}=16/9$, with the value of the scaling exponent decreasing with increasing $n_y$. These estimates are indicated by the dashed lines in Fig.~\ref{fig:aKZvstrap}, and are clearly inconsistent with the measured values.

\subsection{Position-dependent vortex suppression}

\begin{figure}[t]
    
    \includegraphics[width=3.2in]{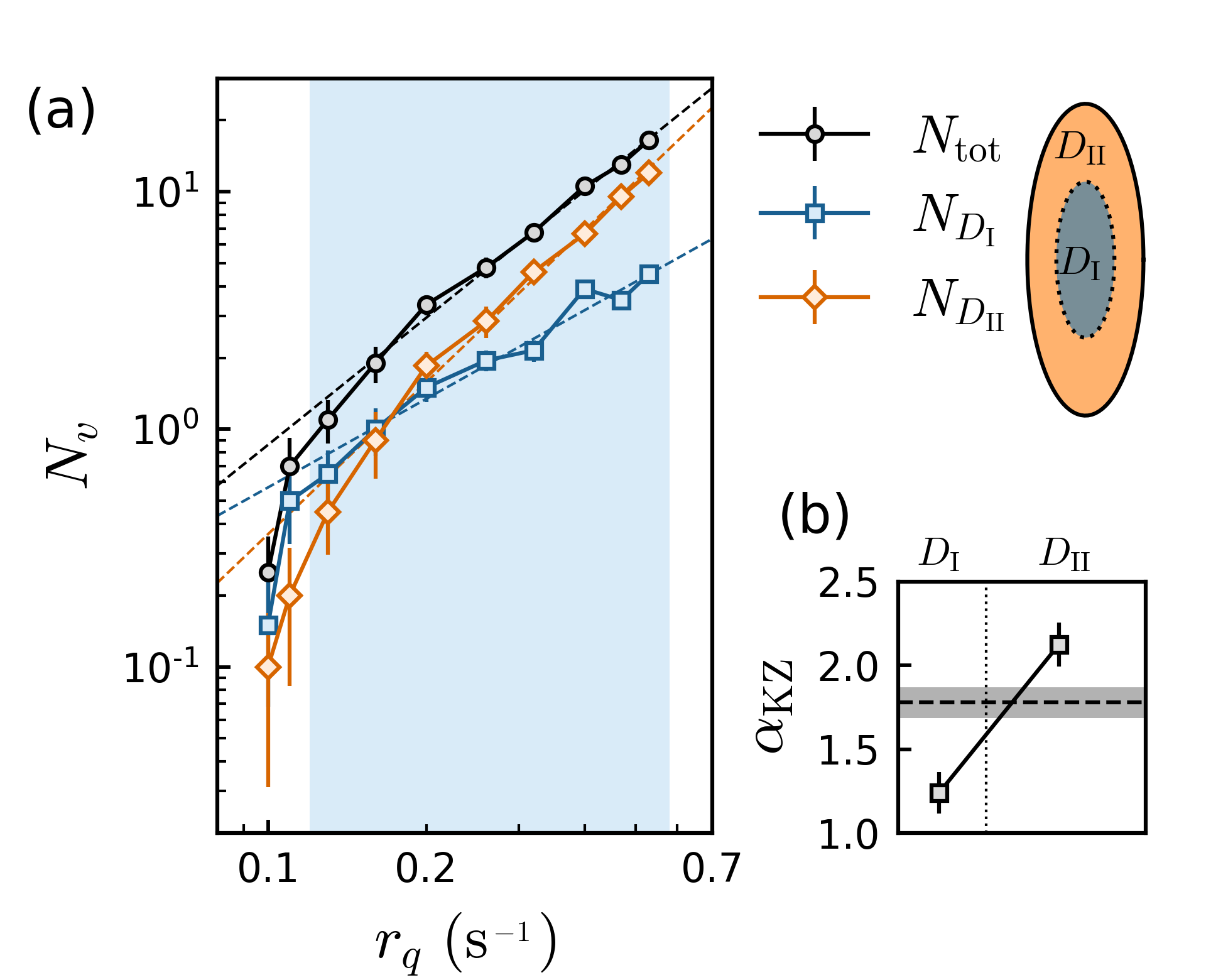}%
    
    \caption{Vortex position analysis in trap D. The sample is divided into two regions: an inner region $D_\text{I}$, and an outer region $D_\text{II}$. (a) The vortex number in each region is plotted against the quench rate $r_{q}$. The blue and red curves represent the vortex numbers in region $D_\text{I}$ and $D_\text{II}$ respectively. The black curve represents the total vortex number [the same data as in Fig.~\ref{fig:allquench}(a)]. The shaded area indicates the KZ scaling region. (b) Scaling exponents in the two regions. The dashed line represents the scaling exponent obtained for the entire condensate area and the gray band is the corresponding $1\sigma$ uncertainty bound.}
    \label{fig:vorvspos}
\end{figure}

As detailed in Ref.~\cite{kim22}, the causality effect, due to the competition between the speed of the spread of the phase transition front and the local speed of sound, leads to a suppression of vortices in the outer regions of a condensate in a power-law trap~\cite{del_Campo_2011,del_Campo_2013}. To study the impact of this effect on vortex suppression in a sample in a harmonic trap, we examine the vortex position distribution in samples created in trap D. As shown in Fig.~\ref{fig:vorvspos}(a), we analyze the quench data by dividing the sample into two regions, an inner region $D_\text{I}$ and an outer region $D_\text{II}$, separated by an iso-density curve, indicated by the dotted line. We fit the data in the scaling region, indicated by the blue-shaded region, by a power-law fit function and extract the relevant scaling exponents. As shown in Fig.~\ref{fig:vorvspos}(b), the scaling exponent increases from 1.2(1) in region $D_\text{I}$, to 2.1(1) in region $D_\text{II}$. This is consistent with the trend observed in trap C~\cite{kim22}, and with the presumed role of the causality effect in vortex suppression in such inhomogeneous systems. Therefore, this further validates the significance of causal interactions for defect formations in systems where different parts undergo phase transitions non-simultaneously. 
 
Nevertheless, we contrast this qualitative agreement between the observations of position-dependent vortex suppression and the predictions of the causality effect, against the aforementioned disagreement in the trend in the KZ scaling exponent for the different trap configurations. These observations unequivocally point toward the importance of the underlying trapping potential geometry for a quantitative understanding of the vortex formation dynamics. 

\subsection{Two-step quench experiments}
\label{sec:3c}

Possible explanations for this discrepancy should entail vortex suppression mechanisms that are more pronounced in certain trap geometries than in others. In this section, we describe an extension of the study of the early-coarsening effect~\cite{Gooetal22} to our harmonic trap D to examine any trap configuration-dependent coarsening during the initial stages of condensate growth. 

\begin{figure*}[htb]
    \includegraphics[width=7in]{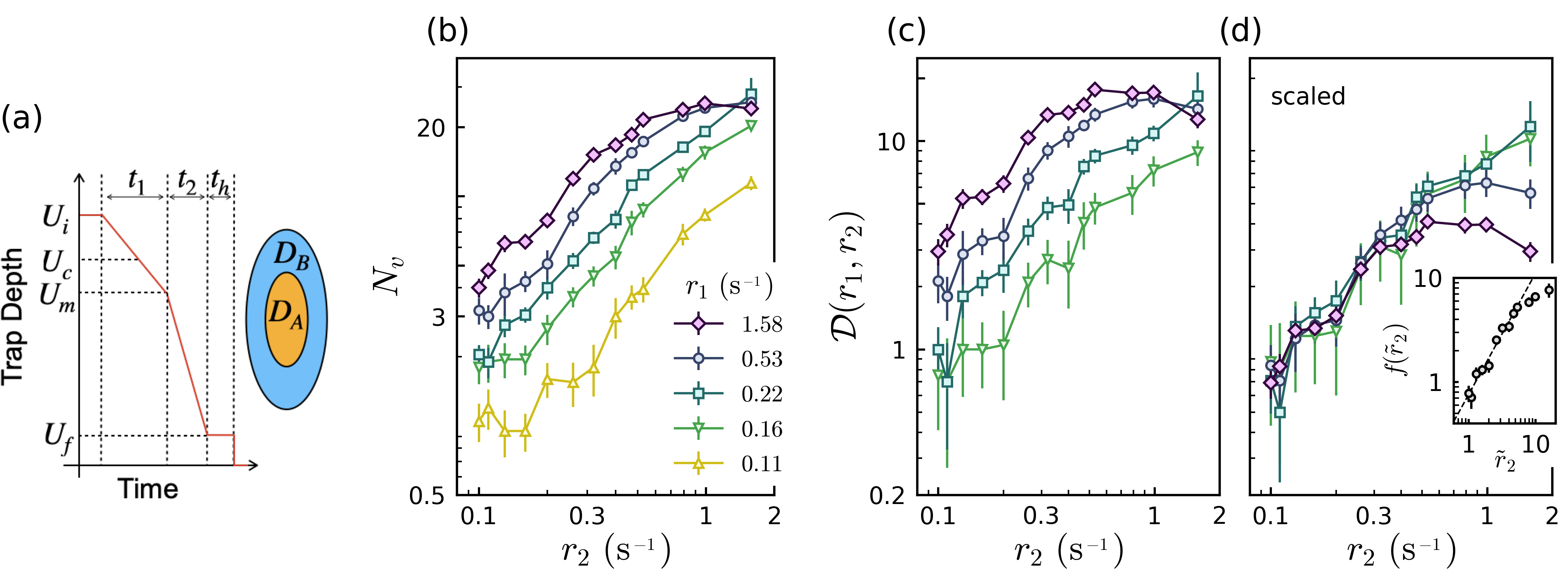}%

    \caption{Two-step quench experiment in trap D. (a) Trap depth evolution for the two-step quench (see the text for details). $D_A$ ($D_B$) represents the inner (outer) region  that undergoes a phase transition for $U>U_m$ ($U<U_m$). (b) Log-log plot of the vortex number $N_{v}$ as a function of the quench rate $r_{2}$ for the second quench step for different former quench rates $r_{1}$. Each data point is the average of 20 realizations of the same experiment and its error bar represents the standard error of the mean. (c) Difference function $\mathcal{D}=N_v(r_1, r_2)-N_v(r_m, r_2)$ with $r_m=0.11~\text{s}^{-1}$ as a function of $r_{2}$. The markers indicate the same value of $r_1$ as in (b). (d) $\mathcal{D}(r_{1},r_{2})$ scaled by the mean of its values at the four lowest $r_{2}$. The inset shows $f(\tilde{r}_{2})$ obtained by averaging the four scaled curves, where $\tilde{r}_2=r_2/r_m$. A power-law fit (dashed line) for the region of $\tilde{r}_2<6$ gives a scaling exponent of 1.16(7).}
    \label{fig:2stepquench}%
\end{figure*}

The Kibble-Zurek description of the vortex formation dynamics rests upon the adiabatic-impulse approximation. As the system approaches the critical point and when the relaxation time exceeds the time it takes to reach the critical point, the spatial correlation of the system essentially freezes. This is referred to as critical slowing down. After passing the critical point and resuming adiabatic evolution, e.g., at $t = t_{\text{fr}}$, the average length scale of domains with correlated choices of the broken-symmetry phase is set by the ``frozen-in'' correlation length. However, here it is proposed that the system only exhibits well-defined domains of independent phases at a later time $t_{d} > t_{\text{fr}}$, allowing for further coarsening of the correlation length during an early stage of the condensate growth defined by $t_{\text{fr}} < t < t_{d}$~\cite{Gooetal22,chesler15}. This is corroborated by the recent observation of a latency time at which the condensate growth commences after the system has crossed the critical point in a harmonically trapped Bose gas~\cite{wolswijk22}. We refer to this as the early-coarsening effect, a key departure from a purely KZ theory governing such vortex nucleation. Traditionally, the power-law scaling exponent $\alpha^{}_{\text{KZ}}$ has been solely associated with the initial ``seeding" process as the system passes through the critical region. Such an early coarsening of the order parameter inevitably alters this initial phase structure and the resultant defect density. This can therefore lead to an overall scaling exponent quantitatively different from the IKZM estimate.

Following the methodology outlined in Ref.~\cite{Gooetal22} for studying the early-coarsening effect in trap C, we replicate the study to assess the role of this effect in vortex formations in trap D. For this, we employ a two-step quench protocol as shown in Fig.~\ref{fig:2stepquench}(a). The thermal cloud is initially quenched from a trap depth of $U_{i} = 1.15U_{c}$ to $U_{m} = 0.8U_{c}$ in time $t_{1}$ at a rate given by $r_{1} = \frac{U_{i}-U_{m}}{U_{c}}\times\frac{1}{t_{1}}$. This is followed by another quench to the final trap depth $U_{f} = 0.36U_{c}$ in time $t_{2}$ at a rate $r_{2} = \frac{U_{m}-U_{f}}{U_{c}}\times\frac{1}{t_{2}}$. Following a hold time of $t_{h} = 1.25\,\text{s}$, the resultant condensate is imaged after a time of flight of 40.9 ms.

We measure the average vortex number $N_{v}$ at different values of $r_{2}$ for a given $r_{1}$, as shown in Fig.~\ref{fig:2stepquench}(b). In an inhomogeneous sample, the density inhomogeneity results in a position-dependent critical temperature, with a higher critical temperature at the trap center compared to the outer regions. Consequently, after the initial quench step, a central sub-region $D_\text{A}$ of the sample undergoes phase transition first, as shown in the inset of Fig.~\ref{fig:2stepquench}(a).  The outer region $D_\text{B}$ undergoes phase transition during the second quench step. As a result, we express the resultant vortex number as a sum of the vortices generated in the two regions: $N_{v} = N_{A}(r_{1},r_{2}) + N_{B}(r_{2})$, where we assume that $N_{B}$ is independent of $r_{1}$.

Within this two-step quench protocol, the early-coarsening effect can be investigated by studying the dependence of $N_{A}$ on $r_{2}$~\cite{Gooetal22}. We isolate the contribution of $N_{A}$ to $N_{v}$ as 
\begin{align}
\begin{split}
    \label{eq:D}
    \mathcal{D}(r_{1},r_{2}) & =  N_{v}(r_{1},r_{2}) - N_{v}(r_{m},r_{2})\\
                    & =  N_{A}(r_{1},r_{2}) - N_{A}(r_{m},r_{2})
\end{split}
\end{align}
where $r_{m} = 0.11\,\text{s}^{-1}$ is the slowest quench rate realized in our experiment. In Fig.~\ref{fig:2stepquench}(c), we plot the difference $\mathcal{D}$ as a function of $r_{2}$ for five different values of $r_{1}$. As seen in Fig.~\ref{fig:2stepquench}(d), upon rescaling this difference by the mean value for the four lowest rates, we note that all the four curves overlap for $r_2<0.5\,s^{-1}$. This justifies a factorization of the form $\mathcal{D} = f(\tilde{r}_{2})\mathcal{D}(r_{1},r_{m})$ with $\tilde{r}_{2} = r_{2}/r_{m}$. It is reasonable then to assume the relation: $N_{A}(r_{1},r_{2}) = f(\tilde{r}_{2})N_{A}(r_{1},r_{m})$. From this we interpret $f(\tilde{r}_{2})$ as the suppression factor that accounts for the reduced vortex number in the central region $D_\text{A}$ due to the coarsening that occurs during the early stages of the order parameter growth.

The inset of Fig.~\ref{fig:2stepquench}(d) shows the vortex suppression function $f(\tilde{r}_{2})$ determined by averaging the values for the five curves. As was observed in trap C~\cite{Gooetal22}, in trap D, $f(\tilde{r}_{2})$ also exhibits a power-law dependence on $\tilde{r}_{2}$ of the form $f(\tilde{r}_{2}) \propto \tilde{r}_{2}^{\beta_{2}}$. Surprisingly, the resultant scaling exponent $\beta_2=1.16(7)$ is also consistent with the measured value of $\beta_2=1.3(2)$ in trap C. The early-coarsening effect, quantified in this manner, shows no significant trap configuration dependence, and therefore does not address the puzzling variation of the overall KZ scaling exponent with the underlying trap configuration. Nevertheless, given these observations, it is certainly worth investigating the universality of the early-coarsening dynamics as quantified above, particularly in the homogeneous density regime. 

On the other hand, we note that the measured value of $\beta_2$ is surprisingly large and comparable to the overall KZ scaling exponent $\alpha^{}_{\text{KZ}} = 1.8(1)$. It is therefore unclear as to how $\beta_2$ and $\alpha_{\text{KZ}}$ relate to each other. Here, it is also worth questioning the validity of the key assumption inherent in our current analysis; by setting $N_\text{B}$ independent of $r_1$, we ignore the role of any causal interactions at the boundary of the initial seed condensate in region $D_\text{A}$ in the growth of condensate in region $D_\text{B}$. In inhomogeneous systems, where presumably both the causality effect and the early-coarsening dynamics play a significant role in vortex suppression during the early stages of condensate growth, it is not so clear how one would disentangle their effects. This further justifies the need for similar investigations in a homogeneous system, where the entire sample undergoes phase transition simultaneously and the role of causally driven vortex suppression can be minimized. 

\section{Conclusion}

We have presented a study of vortex formation dynamics in inhomogeneous samples of Bose gases of rubidium in different trap configurations. The vortex numbers in these samples show the characteristic power-law scaling with the thermal quench rate as predicted by the IKZM. We also observe vortex number saturation at high quench rates, as seen in earlier studies. The analysis of the saturation quench rates and saturated defect densities for various trap configurations suggests that the defect saturation can be understood as a result of the universal dynamics of quenched Bose gases.

We note the qualitative agreement between our observations of position-dependent vortex suppression and the causality effect as incorporated in IKZM. However, we also highlight, quantitative as well as qualitative discrepancies between theoretical predictions for the scaling exponents $\alpha^{}_{\text{KZ}}$ and what is measured in our experiments. These suggest the need for incorporating vortex suppression mechanisms that are trap configuration specific. 

We study one such potential mechanism by extending our prior investigations into the early-time coarsening of the order parameter in an elongated sample (trap C) to our harmonic sample (trap D). With a two-step quench protocol, we find in trap D, as was observed in trap C, that the early-coarsening dynamics can be described by a power-law suppression factor that depends only on the rate of the second quench step. We also note that the measured value of the scaling exponent of the suppression factor is consistent with the value measured in trap C, suggesting that this mechanism is insensitive to the underlying trapping potential geometry. Moreover, we observe that this value is significantly large and comparable to the overall KZ scaling exponent in trap D, reemphasizing the need to incorporate such early-coarsening effect into our understanding of vortex formation dynamics beyond the initial seeding process. It should nevertheless be noted that the above study of the early-coarsening dynamics, based on the division of the condensate into two regions, neglects any causal interactions at their interface and the resultant effects on the coarsening process. Similar studies in the future should probe the validity of this assumption and prescribe ways for incorporating such boundary effects into such an analysis. 

This paper would also benefit from applications to other inhomogeneous systems, as well as to vortex formation studies in homogeneous samples, where the underlying dynamics are significantly different. Most crucially, in homogeneous samples, the role of the causality effect in vortex suppression is greatly reduced, therefore simplifying the interpretation of the observations and their connections to the underlying mechanisms. A direct measurement of the evolution of the correlation length of the condensate at different time scales, similar to what was presented in Ref.~\cite{Navon15}, would also add to these vortex number studies and shed a complimentary light on the underlying coarsening dynamics. 

\begin{acknowledgements}
This work was supported by the National Research Foundation of Korea (Grants No. NRF-2018R1A2B3003373 and No. NRF-2023R1A2C3006565) and the Institute for Basic Science in Korea (Grant No. IBS-R009-D1).
\end{acknowledgements}



\end{document}